\documentstyle[aps]{revtex}

\newcommand{\be}{\begin{equation}}
\newcommand{\ee}{\end{equation}}
\newcommand{\bc}{\begin{center}}
\newcommand{\ec}{\end{center}}
\newcommand{\bea}{\begin{eqnarray}}
\newcommand{\eea}{\end{eqnarray}}
\newcommand{\ba}{\begin{array}}
\newcommand{\ea}{\end{array}}
\newcommand{\nn}{\nonumber}

\def\1{_{1}}
\def\2{_{2}}
\def\3{_{3}}
\def\x{_{\textstyle{x}}}
\def\y{_{\textstyle{y}}}
\def\z{_{\textstyle{z}}}
\def\dd{^{2}}
\def\br{\mbox{\boldmath $r$}}
\def\bro{\mbox{\boldmath $\rho$}}
\def\bk{\mbox{\boldmath $k$}}
\def\bkappa{\mbox{\boldmath $\kappa$}}
\def\bg{\mbox{\bf G}}
\def\cg{\mbox{\boldmath $\cal G$}}
\def\g{\rm {\bf g}}
\def\ms{\mbox{\boldmath $m_{S}$}}
\def\bv{\mbox{\boldmath $V$}}
\def\laplace{\nabla^{2}}
\def\l{\mbox{q}_{\, \mbox{\footnotesize l}}}
\def\la{\mbox{q}_{\, \mbox{\footnotesize l{\scriptsize 1}}}}
\def\lb{\mbox{q}_{\, \mbox{\footnotesize l{\scriptsize 2}}}}
\def\t{\mbox{q}_{\mbox{\small t}}}
\def\ta{\mbox{q}_{\mbox{\small t{\scriptsize 1}}}}
\def\tb{\mbox{q}_{\mbox{\small t{\scriptsize 2}}}}

\begin{document}
\draft
\title{Surface waves at the interface between two viscous fluids}

\author{Arezky H. Rodr\'{\i}guez \thanks{arezky@ff.oc.uh.cu}, J.
Mar\'{\i}n-Antu\~na, H. Rodr\'{\i}guez-Coppola}
\address{Dpto. de F\'{\i}sica Te\'orica, Fac. de F\'{\i}sica, \\
Universidad de la Habana, C. de la Habana, Cuba}

\author{C. Dopazo}
\address{Dpto. de Fluidos, C.P.S, \\
Universidad de Zaragoza, Espa\~na}

\date{\today}
\maketitle

\begin{abstract}
The Surface Green Function Matching analysis (SGFM) is used to study the normal modes
of the interface oscillations between two non-mixed fluids by considering the
difference in their densities and viscosities. The limiting case of viscous-inviscid
system is used for comparison. The role of the viscosity and the density ratios on the
momentum exchange and on the polarization of the surface modes is analyzed.
\end{abstract}


\section{Introduction}

The theory of surface waves in fluids is usually treated using the Orr-Sommerfeld
equation obtained from potential method \cite{landau}. This procedure is useful to find
the characteristics of the wave, such as dispersion relation and damping but becomes
rather complicated when some other features are needed such as polarization and density
of modes.

On the other hand, the inclusion of the viscosities of all of the media make difficult
to understand the physics of the interface. Broadly speaking, the dual effects of
viscosity is well known \cite{landau}: to {\it dissipate} the energy of any
disturbance, but also it has the more complicated effect of {\it diffusing} momentum.
At present, the theory for viscous cases is not nearly as complete or general as for
inviscid cases and it provides only a partial understanding of the role of viscosity in
such systems.

A suitable formalism for including all the viscosities with great ease in
non-homogeneous systems studying the response function has been developed elsewhere
\cite{gm77,vgm79,pvgm81,gmv92}. This formalism, the method of Surface Green Function
Matching (SGFM), has been extensively used to study various inhomogeneous problems
involving surface waves at solid surfaces, both free solid surface (interface between
vacuum and solid) and solid-solid interfaces \cite{vgm77}. It has also been used in
interface involving fluids \cite{gm77,pvgm81} as is the case of solid-fluid interface
and even fluid-fluid interface (this last case analyzed to give an unified treatment of
waves in solids and fluids which seem to be apparently unconnected problems). As far as
we know, there are no previous works where the SGFM have been applied to the
hydrodynamics problems as these authors suggested.

The aim of this paper is to apply the SGFM to the study of the physical characteristics
(dispersion relation, damping and polarization) on the interface normal modes of two
fluids at rest giving insight of the mechanisms of momentum exchange through the
interface for different ratios between the viscosity and density of the two media.

In the next section a brief outline of the main points of the SGFM is given for the
fluid-fluid system at rest, highlighting the considerations made in the solution of the
problem. Section III is devoted to the physical analysis of the polarization of the
modes and the momentum exchange across the interface. In section IV it is carried out a
numerical evaluation considering the physical interpretation of the terms and the
results for pair of fluids which are analyzed as illustration. Finally some conclusions
are outlined.

\section{SGFM for two viscous fluids including surface effects}

The formal development of the SGFM method has been fully explained elsewhere
\cite{gm77,gmv92} and in particular the treatment of matching with discontinuities
\cite{vgm79}, suitable for the case of two non mixed fluids where the interface has
special effects not seen in the liquid bulk. Mathematical and formal details can be
found elsewhere \cite{gm77} and need not be repeated here. It is only necessary to add
that in fluid-fluid interfaces it is better to work with the velocity of the fluid
particle in agreement with the Navier-Stokes equation, instead of the fluid particle
deformation, suitable when solids are present.

Consider a system formed by a fluid $M\1$ for $z<0$ and a fluid $M_{2}$ for $z>0$, both
of them at rest. It has a planar interface at $z=0$. Analysing first each bulk media
individually to prepare its description in a suitable way for the eventual matching at
the interface, the coordinate system will be choosen considering the planes $z=$const
as those of interest. The notation will be for coordinates $\br=(\bro,z)$,
$\bk=(\bkappa,q)$ where
$\rho$ and $\kappa$ are 2D vectors.

As explained in \cite{gmv92}, the SGFM start with the knowledge of the Green function
G.F of the excitation studied in each bulk material constituent. Then, it is needed to
analyze the physical model for the excitation to perform later matching at the
interface.

Now, to know the G.F of each bulk media, the 3D differential equations of hydrodynamics
are the starting point. The fluids are usually treated as incompressible and described
with the Navier-Stokes equation. However, as explained in \cite{vgm79}, it proves
convenient here to give the theory for {\it compressible} fluids, even if
compressibility effects are ultimately neglected. Then, the equation of mass for
isoentropic processes, and the momentum conservation equation that govern the fluid
motion are linearized by neglecting all nonlinear terms in disturbance quantities. They
may be written, respectively, as
\bea
\label{mass}
&& \frac{1}{c\dd}\frac{\partial}{\partial t} p(\br,t) + \rho \nabla\cdot\bv(\br,t) = 0
\\
\label{momentum}
&& \rho \frac{\partial}{\partial t} \bv(\br,t) = - \nabla p(\br,t) + \eta
\laplace\bv(\br,t) + \left( \eta'+ \frac{\eta}{3} \right) \nabla \nabla\cdot\bv(\br,t)
\eea
where $c$, $p$, $\rho$, $\eta$ and $\eta'$ are the velocity of sound, dynamical
pressure, equillibrium density, shear and bulk viscosities respectively, all of them
considered as constants in each medium. $\bv(\br,t)$ is the velocity of the fluid. We
neglected the external forces and supposed that the perturbation is small enough to
neglect the convective term for pressure in (\ref{mass}).

All space and time dependent quantities will be Fourier transformed according to
$exp[i(\bkappa \cdot \bro - \omega t)]$ where $\omega$ is a frequency. Then, for
surface wave propagation, the amplitudes are functions of
$(\bkappa,\omega)$ on one hand and of $z$ on the other. This $z$
dependence is due to the fact that there is no spatial invariance in this direction and
the Fourier transform can not be accomplished. Green functions, including the ones for
the bulk material constituents, are then conveniently expressed as
$\bg(\bkappa,\omega;z,z')$ or, simply, as
$\bg(z,z')$, with $(\bkappa,\omega)$ understood everywhere.

Time Fourier transform will be implied now on. From eq. (\ref{mass}) it is obtained
$p(\br,\omega)=(\rho c\dd/i\omega)\nabla\cdot\bv(\br,\omega)$, which putted in eq.
(\ref{momentum}) gives rise to
\be
i\rho\omega V_{i}(\br,\omega) + (\bar\Gamma - \eta) \frac{\partial}{\partial x_{i}}
\nabla\cdot\bv(\br,\omega) + \eta \laplace V_{i}(\br,\omega) = 0
\ee
with $i=x,y,z$ and
\be
\label{Gamma}
\bar\Gamma = - \frac{\rho c\dd}{i\omega} + \left(\eta' + \frac{4}{3} \eta
\right)
\ee
as the system of equations which couples the velocity components. This system must be
solved as a whole as it can not be decoupled in the general case.

The actual $\bg(z,z')$ of each bulk media considered separately as infinity can be
obtained in different ways but using, for instance, the Fourier transform 3D, it yields
for the G.F \cite{gmv92}:
\be
\label{5}
\bg (\bk,\omega)=\frac{1}{i\rho\omega - \eta k\dd}
\left [ \mbox{\bf I} + \frac{(\bar\Gamma - \eta) \bk
\bk}{i\rho\omega - \bar\Gamma k\dd} \right ]
\ee
where {\bf I} is the unit matrix and $\bk \bk$ is a diadic product of the wave vector.

Its poles
\be
\label{2.9}
\l=\left( \frac{i \rho \omega}{\bar\Gamma} - \kappa\dd \right)^{1/2} \hspace{0.5cm}
\t=\left( \frac{i \rho \omega}{\eta} - \kappa\dd \right)^{1/2}
\ee
describe the transverse and longitudinal modes of the infinite medium. In (\ref{5}) the
incompressible fluid can be considered taking
$(\bar\Gamma \rightarrow \infty)$ and the proper limit is achieved.

There is no physical reason for the preference of a particular direction in the
$xy$-plane. This spatial symmetry of the system allows us to define, for instance,
$\bk=(0,\kappa,q)$ without loosing generality but getting simplification of the
calculations.

Note that $\l \rightarrow i |\kappa|$ if the compressibility is neglected, see eq.
(\ref{Gamma}), given rise to a vanishing longitudinal mode. So, the
$\l$ pole describes the longitudinal mode due to the compressibility of the
media.

Let $\bg_{S}$ be the Green function (G.F) of the surface system just defined and
$\cg_{S}$ its surface projection. Let $\cg_{S}^{-1}$ be the reciprocal of
$\cg_{S}$ in the two-dimensional $\rho$ or $\kappa$ space. This is the
central object in the SGFM analysis. In particular, knowing $\cg_{S}^{-1}$ it is
possible to find the surface mode dispersion relation (SMDR) and the density of modes
of the surface system \cite{gm77}. It is important to stress that the secular equation
for the SMDR, namely
\be
\label{2.1}
\det \; \cg^{-1}_{S} = 0
\ee
expresses the continuity of the velocity and the stress components transmitted across
$z=0$. This is where the physics of the surface effects comes into the picture.  These
effects introduce changes in the stress components transmitted across the interface and
are ultimately measured by some surface tensor $\ms$ whose physical meaning is that
$\ms$, acting on the velocity field $\bv$, yields the extra forces per unit area
transmitted across the interface.

Let us call $\cg_{SO}^{-1}$ to $\cg_{S}^{-1}$ in the absence of such surface effects,
then one finds \cite{vgm79}
\be
\label{2}
\cg^{-1}_{S} = \cg^{-1}_{SO} + \ms
\ee

Thus the problem is to find $\ms$ for the surface effects one wishes to study. It will
be included in this case only the surface tension $\gamma$ according to Laplace's Law.
It can be deduced \cite{vgm79} that
\be
\label{ms}
\ms = \left\Vert \begin{array}{ccc}
 0 & 0 & 0 \\
 0 & 0 & 0 \\
 0 & 0 & - \frac{\textstyle{\gamma \kappa\dd}}{\textstyle{i \; \omega}} \\
\end{array} \right\Vert
\ee

There is a little difference between the former expression and the expression obtained
in \cite{vgm79} according to the fact that here the velocity of the fluid particle is
considered instead the fluid particle deformation.

Then, eq. (\ref{2}) expresses the continuity of the velocities and the stress
components transmitted across the interface at $z=0$. Knowing
$\cg^{-1}_{S}$ one can find the dispersion relation of the surface modes
(SMDR) through the secular equation (\ref{2.1})

\section{Physics and polarization of the surface modes}

The construction of $\cg^{-1}_{SO}$ is explained in \cite{gm77}. The result, after
adding (\ref{ms}), is
\be
\label{3.1}
\cg_{S}^{-1} = \left\Vert \begin{array}{ccc} \eta\1 \ta + \eta\2 \tb  & \mbox{\boldmath
$0$} \\ \mbox{\boldmath $0$} & \Vert \g_{S}^{-1} \Vert  \\
\end{array} \right\Vert
\ee
where $\g_{S}^{-1}$ is a $2\times 2$ matrix and \mbox{\boldmath $0$} is the null vector
$1\times 2$. $\g_{S}^{-1}$ has components
\bea
\label{3.2.1}
(\g_{S}^{-1})_{11} & = & \frac{\rho\1 \omega \la}{\kappa\dd + \la \ta} + \frac{\rho\2
\omega \lb}{\kappa\dd + \lb \tb} \\
\label{3.3}
(\g_{S}^{-1})_{22} & = & \frac{\rho\1 \omega \ta}{\kappa\dd + \la \ta} + \frac{\rho\2
\omega \tb}{\kappa\dd + \lb \tb} - \frac{\gamma \kappa\dd}{i \; \omega} \\
\label{3.4}
(\g_{S}^{-1})_{12} & = & - (\g_{S}^{-1})_{21} =
\left( \frac{\rho\1 \omega \kappa}{\kappa\dd + \la \ta} + 2 i \kappa
\eta\1 \right) - \left( \frac{\rho\2 \omega \kappa }{\kappa\dd + \lb \tb} + 2 i \kappa
\eta\2 \right)
\eea

We shall refer to the modes as {\it sagittal} or S polarized with
$(0,V\y,V\z)$, {\it transverse tangent} or TT$(V\x,0,0)$, {\it
longitudinal} or L$(0,V\y,0)$ and {\it transverse normal} or TN$(0,0,V\z)$ modes,
according to the component of the velocity they have.

Now, on using (\ref{3.1}) in (\ref{2.1}) the factorisation of the
$(\cg_{S}^{-1})_{11}$ matrix element yields a TT mode which does not
interact with the others, whose dispersion relation is
\be
\eta\1 \ta + \eta\2 \tb = 0
\ee
and has $x$-axis polarization.

It is easily seen according to (\ref{2.9}) that the TT mode has no solution but as
stressed in \cite{pvgm81}, it does contribute to the density of modes and therefore
plays a non trivial role in the physical properties of the interface. This mode exists
but it is not a stationary one if there is other surface effects considered
\cite{pvgm81}.

The rest of (\ref{3.1}) yields the secular equation
\be
\label{3.5}
\det \g_{S}^{-1} = 0
\ee
It gives a sagittal mode with polarization S$(0,V\y,V\z)$ and surface tension included.
It will be analyzed in the following.

The factor $(\g_{S}^{-1})_{11}$, see eq. (\ref{3.2.1}), represents the surface movement
component in $y$ direction due to compressibility of the media while
$(\g_{S}^{-1})_{22}$ is a $z$ direction surface movement. The factor
$(\g_{S}^{-1})_{12}$ represents a coupling between $y$ and $z$
movements giving rise to an S polarization mode. It means that the surface has both
horizontal and vertical movements. In other words, the surface particles move in a kind
of circular orbits depending of its phase difference.

On the other hand, there are no important velocities in our system, then
compressibility can be neglected as described in \cite{gm77} and we will discuss
whether the S polarization remains or not. Putting $\l = i |\kappa|$ in (\ref{3.2.1}),
(\ref{3.3}) and (\ref{3.4}) it is obtained
\bea
\label{3.6}
(\g_{S}^{-1})_{11} & = & \frac{\rho\1 \omega i |\kappa|}{\kappa\dd + i |\kappa| \ta} +
\frac{\rho\2 \omega i |\kappa|}{\kappa\dd + i |\kappa| \tb} \\
\label{3.7}
(\g_{S}^{-1})_{22} & = & \frac{\rho\1 \omega \ta}{\kappa\dd + i |\kappa| \ta} +
\frac{\rho\2 \omega \tb}{\kappa\dd + i |\kappa| \tb} - \frac{\gamma \kappa\dd}{i \;
\omega} \\
\label{3.8}
(\g_{S}^{-1})_{12} & = & - (\g_{S}^{-1})_{21} = \left(\frac{\rho\1 \omega
\kappa}{\kappa\dd + i |\kappa| \ta} + 2 i \kappa \eta\1 \right) -
\left( \frac{\rho\2 \omega \kappa}{\kappa\dd + i |\kappa| \tb} + 2 i \kappa \eta\2 \right)
\eea

First of all let us consider the special case where the viscosity of $M\1$ is
neglected. If we put $\eta\1=0$ in (\ref{3.6})-(\ref{3.8}) it is obtained
$q_{\mbox{\small t{\scriptsize 1}}}\rightarrow\infty$ and hence
\be
\label{3.9}
\g_{S}^{-1} = \left\Vert \ba{cc} \frac{\textstyle{\rho_{2} \omega i
|\kappa|}}{\textstyle{\kappa^{2} + i |\kappa| \tb}} & - \left( \frac{\textstyle{\rho\2
\omega \kappa}}{\textstyle{\kappa\dd + i |\kappa| \tb}} + 2 i \kappa \eta\2 \right) \\
\left( \frac{\textstyle{\rho\2 \omega \kappa }}{\textstyle{\kappa\dd + i
|\kappa| \tb}} + 2 i \kappa \eta\2 \right) & \frac{\textstyle{\rho\1
\omega}}{\textstyle{i \; |\kappa|}} + \frac{\textstyle{\rho\2 \omega
\tb}}{\textstyle{\kappa^{2} + i |\kappa| \tb}} - \frac{\textstyle{\gamma
\kappa^{2}}}{\textstyle{i \; \omega}} \ea \right\Vert
\ee

Note that even though the viscosity of one of the constitutient media was neglected,
the coupled factor remains due to the nonzero viscosity of the other fluid. So, in this
limit this mode remains as sagittal S$(0,V\y,V\z)$ exhibiting movements in
$y$- and $z$-axis for the
surface particles. The fluid was taken as incompressible but there is still a component
of velocity on $y$-direction. As far as we know, nobody has ever pointed out this fact
clear, except Lucassen in his works \cite{llr67,l68,lrl69,lvdt72}, who considered
incompressible fluids, but the movement in $y$-axis was due to active materials on the
surface, no as an effect of viscosity. This coupling of movements could be responsible
for a less wavelength and a bigger dissipation as it will be seen later. It is in this
direction where the viscosity plays an important role.

There is more information in eqs. (\ref{3.6})-(\ref{3.8}). If viscosities are neglected
$(\eta\1=\eta\2=0)$ in $(\g_{S}^{-1})_{11}$ and
$(\g_{S}^{-1})_{12}$ these expressions vanish because $\ta \rightarrow
\infty$, $\tb \rightarrow \infty$, but doing the same in
$(\g_{S}^{-1})_{22}$
does not yield a vanishing result. This leads to:
\be
\label{noviscous}
\g_{S}^{-1} = \left\Vert \ba{cc}
 0 & 0 \\
 0 & \rho\1 \omega\dd + \rho\2 \omega\dd - \gamma \kappa\dd |\kappa|
\ea \right\Vert
\ee

The non zero component of Eq. (\ref{noviscous}) is a generalization of the Kelvin
dispersion relation when the density of the upper medium is included.

Then, it is concluded that when there is no viscosities the S polarized mode becomes TN
mode because the coupled factor disappears and only remains $(\g^{-1}_{S})_{22}$.
Indeed, if at least one of the viscosities is considered the coupled factor appears
giving rise to the S mode, (see eq. (\ref{3.9})). Hence, the viscosity is the
fundamental force which couples different modes among them.

Furthermore, the viscosity is the main cause of momentum exchange between the two media
through the surface on the $y$-direction movement. Note that the longitudinal component
movement disappears according to $(\g_{S}^{-1})_{11} \rightarrow 0$ when $\eta\1$ and
$\eta\2$ are neglected. On the other hand, in (\ref{3.5}) the transverse
normal movement described by $(\g_{S}^{-1})_{22}$ exists because of the densities and
viscosities of the media, (see eq. (\ref{3.7})). When the viscosities are neglected as
in (\ref{noviscous}), the normal component movement still remains because of the
densities of the fluids. Hence, in the case of the $z$-axis movement both the
viscosities and densities are important for the exchange of momentum.

These results are in agreement to the fact that when the interface particle moves
according to the longitudinal mode it remains on the plane
$z=0$ and the viscosities are the only way for the two media to interact,
but when the interface particle moves according to the transverse normal mode it goes
into each medium sometimes at $z>0$ and other at $z<0$ and then the inertial effects of
the media become important according to their densities.

Expressions (\ref{3.5})-(\ref{3.8}) also recover the Kelvin equation for an ideal fluid
with free surface, (see references in \cite{lrl69,landau}). Neglecting the viscosities,
and setting $\rho\1=0$ it is obtained
\be
\g_{S}^{-1} = \left\Vert \ba{cc} 0 & 0 \\ 0 &  \rho\2 \omega\dd - \gamma \kappa\dd
|\kappa| \\ \ea \right\Vert
\ee

It can be seen that our formalism not only recovers the expression for the Kelvin
equation, but also recovers the $z$-polarization of that mode.

After this analysis one can return back to the problem for both viscous fluids. From
(\ref{3.5})-(\ref{3.8}) it is obtained the secular equation for the SMDR
$$
\omega^2 \left[(\rho\1 + \rho\2) (\rho\1 \tb + \rho\2 \ta) - |\kappa| (\rho\1 -
\rho\2)\dd \right] + \gamma \kappa\dd |\kappa| \left[ \rho\1 (|\kappa| - \tb) + \rho\2
(|\kappa| - \ta) \right] +
$$
\be
\label{2.4}
+ 4 \kappa\dd |\kappa| (\eta\2 - \eta\1)\dd(|\kappa| - \ta)(|\kappa| - \tb) + 4 i
\kappa\dd \omega (\eta\2 - \eta\1)(\rho\1 |\kappa| - \rho\2 |\kappa| - \rho\1 \tb +
\rho\2 \ta) = 0
\ee
with the following new definition of $\t=(\kappa\dd - i \rho \omega/\eta)^{1/2}$. This
expression, which we recall corresponds to two viscous non mixed incompressible fluids,
can also be accomplished applying the potential method, although using that formalism
it is rather difficult to obtain the polarization of the modes.

This is the equation to be used to study the modes if one includes both viscosities and
surface tension effects for incompressible fluids. Expression (\ref{2.4}) was reported
in \cite{pvgm81} to study the surface waves at the interface between a solid and a
fluid. They neglected the surface tension. One of the aim of this paper is to compare
this theory with the theory which just take into account only one of the viscosities.
From expression (\ref{3.9}) it is not difficult to achive the SMDR for the
viscous-inviscid fluid interface
\be
\label{37}
- \omega\dd \rho\2 (\rho\1 + \rho\2) + \gamma \kappa\dd |\kappa| \rho\2 + 4 \kappa\dd
|\kappa| \eta\2\dd (|\kappa| - \tb) - 4 i \rho\2 \omega \kappa\dd \eta\2 = 0
\ee
which reduces to equation (2.5) of \cite{pvgm81} when $\gamma=0$ and will be evaluated
in the next section along with (\ref{2.4}) for the viscous case.

\section{Results of the numerical evaluation}

In order to make a numerical study the following quantities of length and time for
nondimensionalization will be taken:
\bea
\makebox{time by} \hspace*{10mm} T_{O} & = & \frac{\eta^{3}\2}{\rho\2 \gamma\dd}
\nonumber
\\
\makebox{length by} \hspace*{10mm} L_{O} & = & \frac{\eta\2\dd}{\rho\2 \gamma}
\eea
The dispersion relation (\ref{2.4}) becomes
\bea
\label{40}
\omega\dd \left[(1+Q)(\bar{\ta} + Q \bar{\tb}) - |\kappa|(1-Q)\dd
\right] + \kappa\dd |\kappa| \left[|\kappa|(1+Q)-\bar{\ta} - Q\bar{\tb}
\right] & + & \nn \\
+ 4 \kappa\dd |\kappa| (1-N)\dd (|\kappa| -\bar{\ta})(|\kappa| -\bar{\tb}) + 4 i
\kappa\dd \omega (1-N)\left[- |\kappa|(1-Q) + \bar{\ta} - Q \bar{\tb}
\right] & = & 0
\eea
for viscous fluids and eq. (\ref{37}) gives rise to
\be
\label{non-viscous}
- \omega\dd \left(1+Q\right) + \kappa\dd |\kappa| + 4 \kappa\dd |\kappa| (|\kappa|
-\bar{\tb}) - 4 i \omega \kappa\dd = 0
\ee
for viscous-inviscid case, where $Q = \rho\1/\rho\2$, $N=\eta\1/\eta\2$ and
\bea
\bar{\ta} & = & \left( \kappa\dd - i \omega \frac{Q}{N} \right)^{1/2} \\ \bar{\tb} & =
& \left( \kappa\dd - i \omega \right)^{1/2}
\eea

Then the characteristics of the system will be studied by its SMDR with real values of
the frequency $\omega$. Let us allow
$\kappa$ to be complex, its real part is $2\pi$ times the inverse of the
wavelength and the imaginary part is the distance damping coefficient
$\beta$ related with the viscosities of the media. The dimensionless
parameters are $\kappa_{o}=2\pi/L_{o}$ and $\omega_{o}=2\pi/T_{o}$.

Fig. \ref{rdq8} shows the SMDR for $Q=0.8$. There is one mode which decreases its
wavelength $\lambda$ and increases its distance damping coefficient
$\beta$ with increasing frequency at a fixed value of the parameter $N$.
It is also shown that when the viscosity ratio $N$ is increased the wavelength lightly
decreases at any frequency. The curves split bigger at higher frequencies. On the other
hand $\beta$ increases with increasing $N$. Also it was plotted the curves obtained
with $N=0$ from eq. (\ref{non-viscous}) which means zero viscosity of the medium $M\1$.
It can be seen that the theory which includes all the viscosities predicts small
$\lambda$ and bigger $\beta$ for a fixed
$\omega$ with respect to the $N=0$ case.

Fig. \ref{v2qw1} shows the dependence of $\kappa$ and $\beta$ with respect to the
variation of the density ratio $Q$ at a fixed value of the frequency and viscosity
ratio. It can be seen that $\lambda$ decreases when $Q$ increases at a fixed $N$. This
was deduced by Taylor in his study of the ripple formation on an infinitely thick
viscous circular jet but neglecting the air viscosity. References are given in
\cite{llc}. We now prove that this is also true when both viscosities are considered.
Also $\lambda$ decreases at a fixed $Q$ when the viscosity ratio $N$ takes higher
values. So, the effect of the viscosity of medium $M\1$ reinforces the effect produced
by the density and it can be stated that the smaller wavelength will be obtained when
$Q$ and $N$ are both bigger. It is also plotted the curve with
$N=0$ corresponding to the viscous-inviscid case. It can be seen that the
wavelength is always smaller in the case $N\neq0$ (viscous-viscous case). The curves
split bigger as $Q$ increases and $\beta$ grows rapidly at low values of $Q$ for a
fixed value of $N$ and tends to saturation for higher values of $Q$. This small
variation of $\beta$ with the variation of
$Q$ even at a fixed value of the viscosity ratio $N$ reinforces
the idea of the density as another mechanism of momentum exchange between the two media
through the interface. It not only produces smaller wavelengths, but also produces
lightly bigger distance damping coefficients $\beta$.

On the other hand, the distance damping coefficient $\beta$ also increases at higher
values of the viscosity ratio $N$ at a fixed $Q$. It was also plotted the curves at
$N=0$. It is interesting to note that the theory of viscous-inviscid case predicts a
small decrease of the distance damping coefficient with increasing density ratio $Q$.
This is in accordance to the fact that setting $N=0$ means to neglect the momentum
exchange through the interface by the viscosity and raising $Q$ represents to increase
the dynamic properties of the surface given rise to a bigger distance for the wave to
travel before vanishing.

Fig. \ref{v2vw1} shows the variation of $\lambda$ and $\beta$ with respect to the
viscosity ratio for a fixed value of $\omega$ at three values of
$Q$. It shows that as the viscosity ratio increases, the wavelength reduces
rapidly first and tends to a limiting value for $N\geq 1$. The curves start in the
value of $\lambda$ corresponding to the viscous-inviscid case. Also, for a fixed value
of $N$ the wavelength decreases as $Q$ increases, in correspondence with Fig.
\ref{v2qw1}. For the coefficient $\beta$ it is seen that it raises for increasing $N$.

To illustrate this theory for real fluid combinations there will be used three pairs of
fluids: air/water, water/aniline and water/mercury. The parameters of these fluids at
room temperature are: \bc
\begin{tabular}{cccc} \hline
Element & Density  & Viscosity & Surface \\
        &          &           & Tension \\
        &(kg/m$^3$)&($m$Pa s)  & ($m$N/m)\\ \hline
air     &$1.21$    & 0.018     & -       \\ water   &$998$     & 0.890     & 71.99   \\
mercury &$13500$   & 1.526     & 485.48  \\ aniline &$1022$    & 3.847     & 42.12   \\
\hline
\end{tabular}
\ec

Then, for the system air/water it is $Q=0.0012$ and $N=0.0202$. In this case the SMDR
is plotted in Fig. \ref{air-wat}. It can be seen that there is no difference of the
wavelength reported by viscous-viscous and viscous-inviscid cases due to the small
values of the density and viscosity ratios but there is a small increase of $\beta$ for
all frequencies when the air viscosity is considered.

However, the operating conditions in many gas turbine combustors and liquid-propellant
rocket engines are such that the density and viscosity ratios are higher. It could be
so also in water-oil emulsions and other problems where the interface between two
fluids plays an important role. Then, the SMDR for water/aniline ($Q=0.977$, $N=0.231$)
and water/mercury ($Q=0.074$, $N=0.583$) systems were also plotted. In the first case
(water/aniline) the densities are very similar but the viscosity of the aniline is much
bigger than the water viscosity. In the case of water/mercury the viscosities are near
one half one another, but the density of the mercury is much bigger the density of the
water. The SMDR has been plotted in Fig. \ref{anil-wat} and \ref{mer-wat}. Note than in
both cases the wavelength decreases and the distance damping coefficient increases in a
visible way when the viscosity of the medium $M\1$ (water in both cases) is considered.
The difference is bigger at high values of the frequency where the viscosity effects
become important. Then, one can conclude that the inclusion of the viscosities of all
media produces a substantial decrease of the wavelength if the viscosity ratio are big
enough. It gives rise also to a bigger distance damping coefficient for the wave.

\section{Conclusions}

In the present paper the close relationship between the properties of low amplitude
surface waves propagation with the viscosity and density ratios, in a system of two
non-mixed incompressible fluids at rest has been set out. The SGFM method was used to
accomplish the dispersion relation and the full study of wave propagation by varying
different parameters of the media.

It was shown that the viscosity is a fundamental parameter for the coupling of
different modes. It gives rise to an S polarization mode with $y$ and
$z$ components of the movement of the particles on the surface. Also it was
seen that the viscosity is the main force in producing momentum exchange in the
longitudinal mode, but for the transverse normal mode both the viscosity and the
density ratios are important to the momentum exchange.

When considering surface modes, it was shown that only one of them is allowed and its
wavelength is smaller when considering the viscosity of both media for fixed values of
the density ratio. Also it was seen a characteristic variation of the distance damping
coefficient when the viscosity of all media are included. On the other hand the
increasing of the density ratio also reduces the wavelength and produces a lightly
increase of the distance damping coefficient, then this factor is also important in
reducing the wavelength of the surface waves.

In order to see more real situations, three pair of fluids were analyzed and the
importance of taking in consideration all the viscosities was shown.

\section*{Acknowledgements}

We are indebted to Professors Federico Garc\'{\i}a-Moliner and V. R. Velasco for
advices and clever discussions.

This work was partially supported by an Alma Mater grant, University of Havana.

\begin{figure}
\caption{Dispersion relation of the surface mode for $Q = 0.8$. The upper part gives
the wavelength $\kappa/\kappa_{o}$ and the lower part the distance damping coefficient
$\beta/\kappa_{o}$}
\label{rdq8}
\end{figure}

\begin{figure}
\caption{Relation between wavelength and distance coefficient with respect to the
density ratio at a fixed frequency for different values of viscosity ratios. The case
$N = 0$ is the viscous-inviscid case}
\label{v2qw1}
\end{figure}

\begin{figure}
\caption{Relation between wavelength and distance coefficient with respect to the
viscosity ratio at a fixed frequency for different values of density ratios}
\label{v2vw1}
\end{figure}

\begin{figure}
\caption{Dispersion relation of the surface mode for the air/water system. In the
legend it is especified the $M\1$ as left and $M\2$ as right in the combination
$M\1/M\2$, i. e., air/water in this figure.}
\label{air-wat}
\end{figure}

\begin{figure}
\caption{Dispersion relation of the surface mode for the water/aniline system}
\label{anil-wat}
\end{figure}

\begin{figure}
\caption{Dispersion relation of the surface mode for the water/mercury system}
\label{mer-wat}
\end{figure}

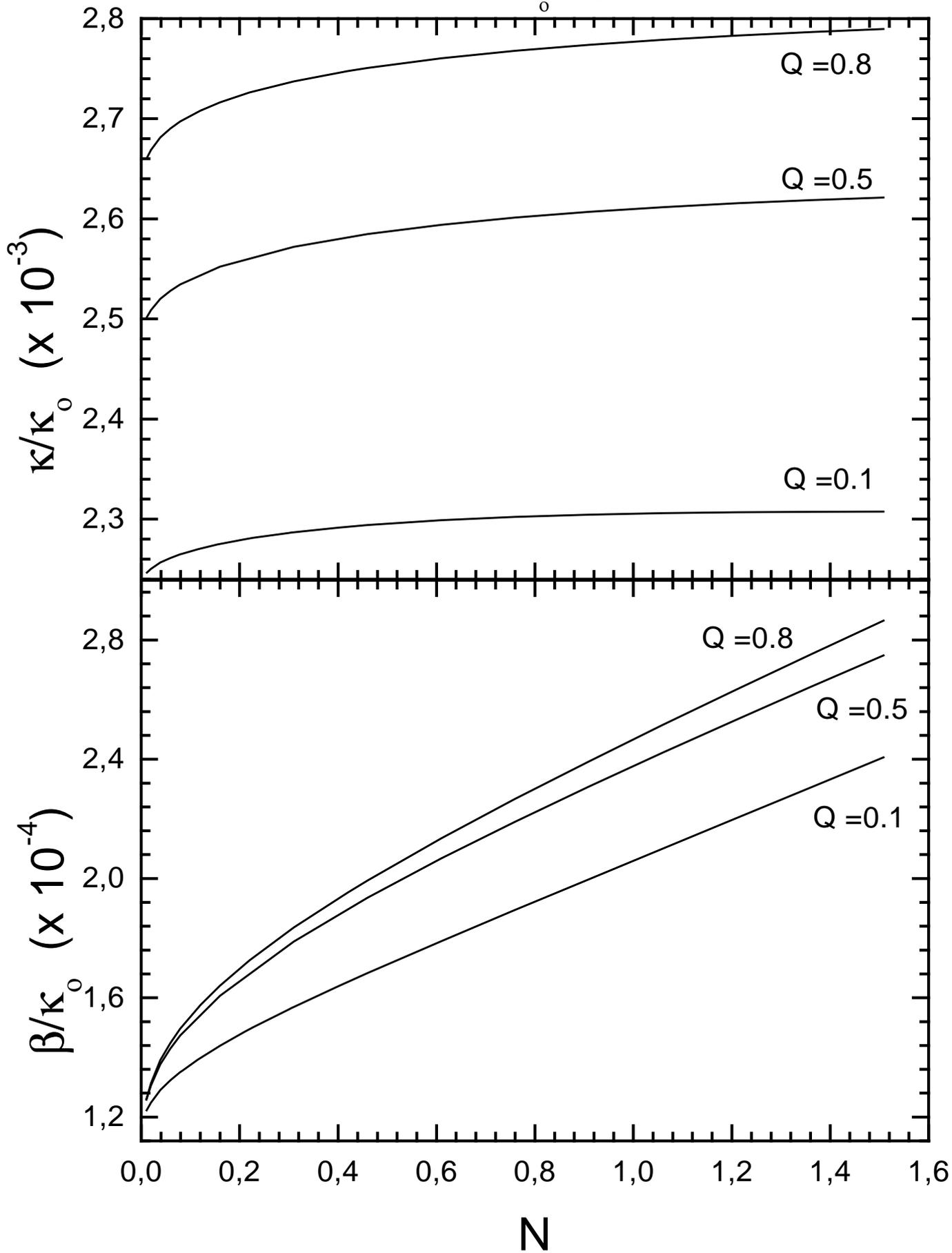
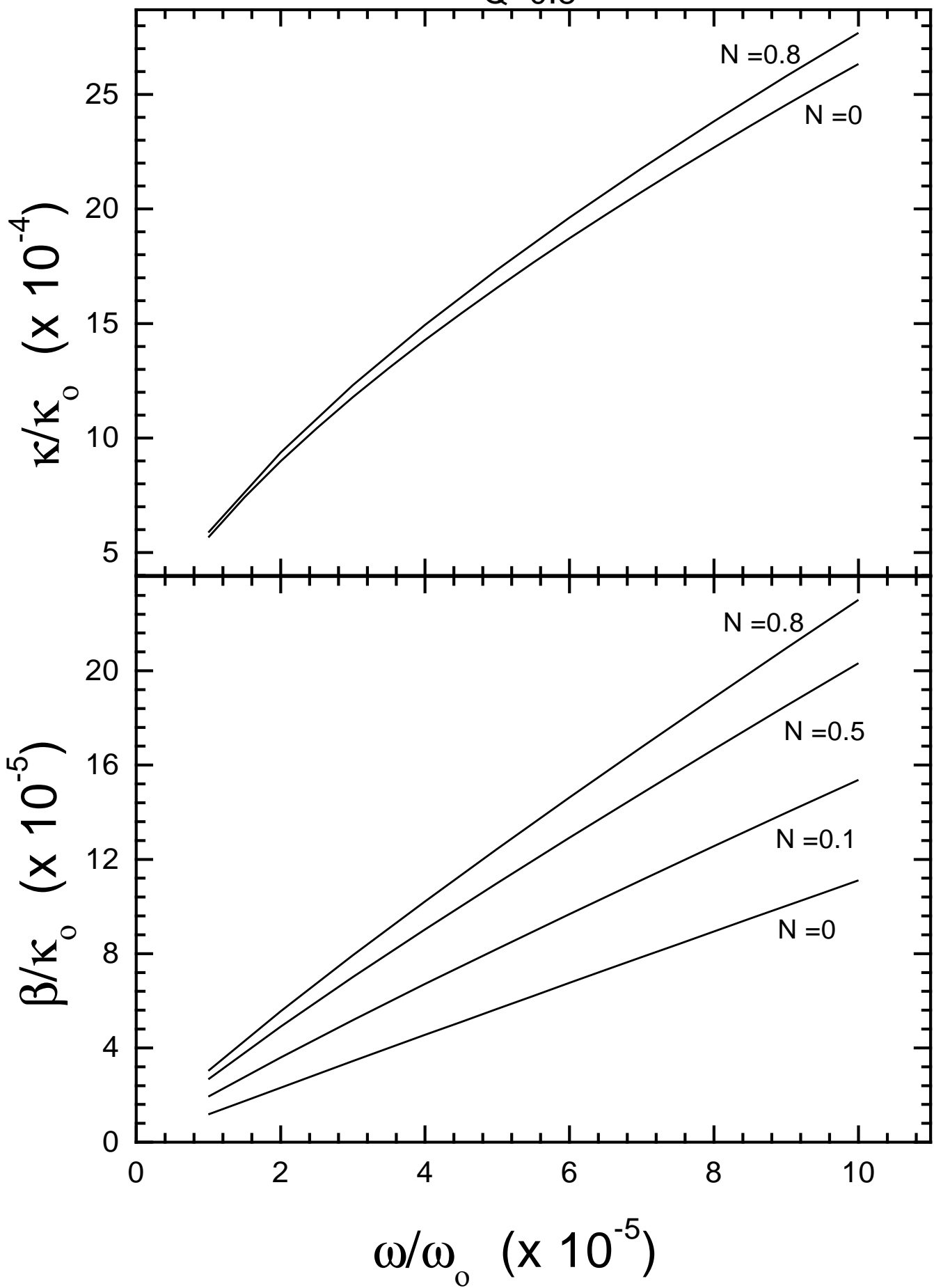
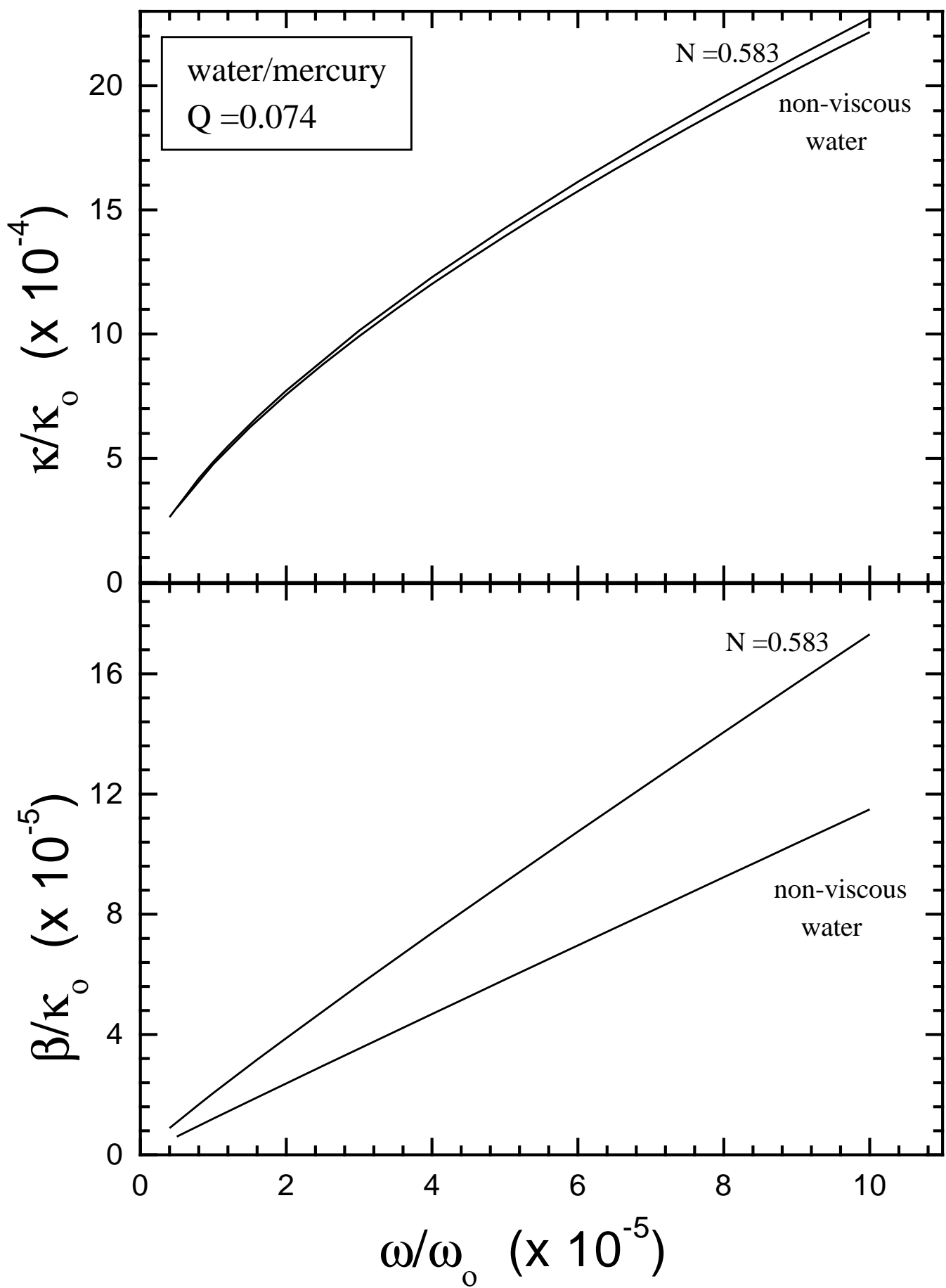
\end{document}